\documentclass[aps,twocolumn,showpacs,amsmath,amssymb]{revtex4}
\usepackage[english]{babel}
\usepackage{graphicx}

\begin{document}
\title{The Wave Function of a Gravitating Shell}
\author{\firstname{V.~I.}~\surname{Dokuchaev}}
\email{dokuchaev@ms2.inr.ac.ru} \affiliation{Institute for Nuclear Research, Russian Academy of Sciences, pr. 60letiya Oktyabrya 7a, Moscow, 117312 Russia}
\author{\firstname{S.~V.}~\surname{Chernov}}
 \email{chernov@lpi.ru}
\affiliation{Lebedev Physical Institute, Russian Academy of Sciences,
Leninskii pr. 53, Moscow, 119991 Russia \\
Institute for Nuclear Research, Russian Academy of Sciences, pr. 60letiya Oktyabrya 7a, Moscow, 117312 Russia}

\begin{abstract}
We have calculated a discrete spectrum and found an exact analytical
solution in the form of Meixner polynomials for the wave function of
a thin gravitating shell in the Reissner-Nordstr\"om geometry. We show
that there is no extreme state in the quantum spectrum of the gravitating
shell, as in the case of extreme black hole.
\end{abstract}


\maketitle

In the absence of a quantum theory of gravity, the semiclassical approximation (gravitons, the Hawking effect, the primordial perturbation spectrum in cosmology) and various semi-qualitative ``plausible'' models (for the black hole mass spectrum) \cite{Much,BekMuch} are used as a convenient method for describing quantum effects against the classical background of general relativity. One of the models, in which it is relatively easy to obtain the black hole mass spectrum, is the model of a thin gravitating shell \cite{IsraelNC}.

There exists a simple method for obtaining the mass spectrum in the
formalism of thin shells. It is based on the natural assumption that
the mass of a gravitating system (e.~g., a black hole) at spatial
infinity, $m_{\rm out}$, is the Hamiltonian of the system, because
$m_{\rm out}$ is the total energy of the entire system that is
conserved during the dynamical evolution of the shell. Given the
Hamiltonian, the wave equation can then be easily written and the
corresponding mass spectrum can be found from its solution.

All properties of a Schwarzschild black hole in general relativity
are known to be completely determined by one parameter, namely, the
mass at spatial infinity. However, if we consider the Carter-Penrose
diagram for an eternal Schwarzschild black hole (see the figure 1),
which describes the global geometry of the manifold in question,
then we will see that there exist two spatial infinities, the
so-called regions  $R_+$ and $R_-$. Therefore, it can be assumed that
the mass spectrum in the Schwarzschild metric of an eternal black
hole can depend on two quantum numbers \cite{BerezinArxiv}.

Consider the dynamical equation that describes the evolution of a
thin gravitating shell in the Reissner-Nordstr\"om metric. This
equation has a well known form (see, e.~g., \cite{BKTPRD87,CherDokRN}
and \cite{DokCherPismaJETP,DokCherJETP,DokCherCQG,BlGuenGuth}):
\begin{eqnarray}
 &&\sigma_{\rm in}\sqrt{\dot{\rho}^2+1-\frac{2m_{\rm in}}{\rho}
 +\frac{Q^2_{\rm in}}{\rho^2}} \nonumber\\
 &&-\sigma_{\rm out}\sqrt{\dot{\rho}^2+1-\frac{2m_{\rm out}}{\rho}
 +\frac{Q^2_{\rm out}}{\rho^2}}
 =4\pi\rho\,\mu(\rho),
 \label{GenEquat}
\end{eqnarray}
where $\sigma_{\rm in,out}=\pm1$, $m_{\rm in}$, $m_{\rm out}$, $Q_{\rm in}$ and $Q_{\rm out}$ are, respectively, the black hole mass and charge inside and out side the shell; $\rho=\rho(\tau)$ is the shell radius relative to an observer on the shell; and the dot over the function denotes a derivative with respect to the proper time $\tau$ of this observer. We will consider only a dust shell, then the function $\mu(\rho)=A/\rho^2$, where $A>0$ is the constant of integration \cite{CherDokRN}. For the subsequent analysis, it is
convenient to define a quantity $M=4\pi A$ that, as can be shown, is the total mass of the shell. As is easy to show from Eq. (\ref{GenEquat}), the conditions for the signs of $\sigma_{\rm in,out}$ are
\begin{eqnarray}
 \label{sigmain}
 \sigma_{\rm in}&=&\mbox{sign}\Bigg[m_{\rm out}-m_{\rm in}
 +\frac{Q^2_{\rm in}-Q^2_{\rm out}+M^2}{2\rho}\Bigg], \\
 \sigma_{\rm out}&=& \mbox{sign}\Bigg[m_{\rm out}-m_{\rm in}
 +\frac{Q^2_{\rm in}-Q^2_{\rm out}-M^2}{2\rho}\Bigg].
 \label{sigmaout}
\end{eqnarray}
Everywhere below, we will consider the case where $m_{\rm out}>m_{\rm in}$, $M^2\ge Q^2_{\rm out}-Q^2_{\rm in}$, and $\sigma_{\rm in}=1$ in the entire spacetime region. At the same time, the sign of $\sigma_{\rm out}$ can have any value. Squaring Eq.~(\ref{GenEquat}) yields an expression of the form
\begin{eqnarray}
 m_{\rm out}&=&m_{\rm in}+\frac{Q_{\rm out}^2-Q^2_{\rm in}-M^2}{2\rho}
 \nonumber\\
 &&+ M\sigma_{\rm in}\sqrt{\dot{\rho}^2+1-
 \frac{2m_{\rm in}}{\rho}+\frac{Q^2_{\rm in}}{\rho^2}}.
 \label{energy}
\end{eqnarray}
The total mass of the entire system $m_{\rm out}$, which is conserved
with time, appears on the left-hand side of this expression. Following
the earlier papers \cite{BerKozKuzTkachPLB,Hajicek,Berezin}, we will
define this mass as the Hamiltonian of our system H (see also
\cite{BerezinPLB}, where a slightly different model was considered).
Let us make the change of variable $x=M\rho$ in Eq.~(\ref{energy})
(see \cite{BerKozKuzTkachPLB,Hajicek,Berezin}). The Hamiltonian will
then take the form
\begin{eqnarray}
 H&=&\sigma_{\rm in}\sqrt{\dot{x}^2+M^2\!\left(\!1-
 \frac{2m_{\rm in}M}{x}+\frac{Q^2_{\rm in}M^2}{x^2}\right)}
 \nonumber\\
 &&+m_{\rm in}+M\frac{Q_{\rm out}^2-Q_{\rm in}^2-M^2}{2x}.
 \label{hamiltonian}
\end{eqnarray}
Given Hamiltonian (\ref{hamiltonian}), we calculate the Lagrangian
of the system under consideration
\begin{eqnarray}
 L\! &=& \!\sigma_{\rm in}\dot{x}\ln\!\!\left[\dot{x}
 +\!\sqrt{\dot{x}^2+M^2\!\left(\!1-
 \frac{2m_{\rm in}M}{x}\!+\!\frac{Q_{\rm in}^2M^2}{x^2}\right)}\right]
 \nonumber\\
 &&-\sigma_{\rm in}\sqrt{\dot{x}^2+M^2\left(1-
 \frac{2m_{\rm in}M}{x}+\frac{Q_{\rm in}^2M^2}{x^2}\right)}
 \nonumber\\
 &&+ M\frac{M^2-Q_{\rm out}^2+Q_{\rm in}^2}{2x}
 -\sigma_{\rm in}\dot{x}\ln M-m_{\rm in},
\end{eqnarray}
 and then the canonical momentum
\begin{equation}
 p\! =\!\sigma_{\rm in}\!\ln\!\!\left[\frac{\dot{x}}{M}\!
 +\!\sqrt{\frac{\dot{x}^2}{M^2}\!+\left(\!\!1\!-\!
 \frac{2m_{\rm in}M}{x}\!
 +\!\frac{Q^2_{\rm in}M^2}{x^2}\!\right)}\,\right].
 \label{impulse}
\end{equation}
\begin{figure}[!t]
\includegraphics[width=0.3\textwidth]{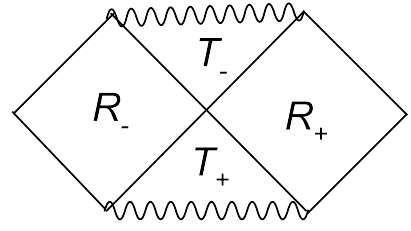}
 \caption{Carter-Penrose diagram for an eternal Schwarzschild black hole.}
\end{figure}
Let us now write Hamiltonian (\ref{hamiltonian}) in terms of the canonical momentum (\ref{impulse}). For this purpose, we will find the variable $\dot{x}$ from Eq.~(\ref{impulse}) and substitute it into Hamiltonian (\ref{hamiltonian}). As a result, the dependence of the Hamiltonian on the coordinate $x$ and canonical momentum $p$ will take the form
\begin{eqnarray}
 H &=&m_{\rm in}-M\frac{M^2-Q_{\rm out}^2
 +Q_{\rm in}^2}{2x} \\
 &&\!+\! \frac{\sigma_{\rm in}M}{2}\!\Bigg[e^{\sigma_{\rm in}p}
 +\!\!\left(\!\!1-\frac{2m_{\rm in}M}{x}+\frac{Q_{\rm in}^2M^2}{x^2}\right)
 \!e^{-\sigma_{\rm in}p}\Bigg]. \nonumber
 \end{eqnarray}
Let us write the wave equation $H\phi(x)=m_{\rm out}\phi(x)$, using a standard
commutation equality of the form $[p,x]=-i$ and the identity
\begin{equation}
 \exp\left(x_0\frac{\partial}{\partial x}\right)\phi(x)=\phi(x+x_0).
\end{equation}
As a result, the wave equation will be written as
\begin{eqnarray}
 \phi(x-i) &+& \left(1-\frac{2m_{\rm in}M}{x}
 +\frac{Q_{\rm in}^2M^2}{x^2}\right)\phi(x+i)
 \nonumber\\
 &-& \frac{M^2-Q^2_{\rm out}+Q^2_{\rm in}}{x}\phi(x)=2E\phi(x),
\label{BasicEquat}
\end{eqnarray}
where $E=(m_{\rm out}-m_{\rm in})/M$. The wave equation (\ref{BasicEquat})
differs from the standard Schrodinger equation in that this is not a
differential equation but a difference one. This is because the
quantization is performed over the proper time of an observer on the shell
and not over the time of an observer at infinity. If we represent the
exponential as a series, then we will obtain an infinite order
differential equation. Consequently, this equation should be
supplemented by infinite boundary conditions. These were found in
\cite{Hajicek} and, in our case, are
\begin{eqnarray}
 \phi^{2l}(0)=0, \quad l=0,1\ldots.
 \label{boundcond}
\end{eqnarray}
It should also be required that the wave functions do not diverge at
spatial infinity.

The solution of the wave equation can be expressed in terms of Meixner
polynomials (see, e.~g., \cite{Hajicek}) that satisfy the equation
\begin{eqnarray}
 &&\sigma(x)[f(x+1)-2f(x)+f(x-1)]
 \nonumber\\
 &&+\tau(x)[f(x+1)-f(x)]+\lambda f(x)=0, \label{MeiEquat}
\end{eqnarray}
where $\sigma(x)=x$, $\sigma(x)+\tau(x)=\mu(\gamma+x)$, and $\mu$ and
$\gamma$ are some constants. We will solve the wave equation
(\ref{BasicEquat}) in an explicit form by reducing it to an equation
of form (\ref{MeiEquat}). To begin with, let us make the coordinate
transformation $x\to -ix$ which means the rotation through $-\pi/2$
in the complex plane. In this case, Eq.~(\ref{BasicEquat}) will take
the form
\begin{eqnarray}
 \phi(x+1) &+& \left(1-\frac{2m_{\rm in}Mi}{x}
 -\frac{Q_{\rm in}^2M^2}{x^2}\right)\phi(x-1)
 \nonumber\\
 &-& i\frac{M^2-Q_{\rm out}^2+Q_{\rm in}^2}{x}\phi(x)=2E\phi(x).
\label{WaveRealEquat}
\end{eqnarray}
First, consider the simple case where $m_{\rm in}=Q_{\rm
in}=0$. The solution of Eq. (\ref{WaveRealEquat}) can then be expressed
in terms of Meixner polynomials and is
\begin{equation}
 \phi_n(x)=C(x)\frac{\beta^x x}{\beta^{2x+2n}}\triangle^n
 \bigg[\frac{\beta^{2x}\Gamma(x)}{\Gamma(x+1-n)}\bigg],
\label{solution}
\end{equation}
where
\begin{eqnarray}
 \beta &=& E+\sqrt{E^2-1},\quad \triangle f(x)=f(x+1)-f(x),\nonumber\\
 C(x) &=& C(x+1),
 \label{beta}
\end{eqnarray}
$\Gamma(x)$ is the gamma function, and $C(x)$ is a periodic function with
a period of 1. Let us expand the periodic function C(x) into a Fourier series:
\begin{eqnarray}
 C(x)=\sum_{k=-\infty}^{\infty}c_k\exp(2\pi ikx).
\end{eqnarray}
Exactly the same factor appears in the solution of the wave equation if we
pass from the coordinate representation to the momentum one
\cite{Berezin,BerezinPRD}. The coefficients $c_k$ can be found from the
boundary conditions to the wave equation (\ref{boundcond}). It is easy to
show that the wave functions $\phi_n(x)$ are orthogonal in the following
sense. If we set $x_i=x$ and $x_{i+1}=x_i+1$, then we can write the
following sum using the properties of Meixner polynomials:
\begin{equation}
 \sum_{x_i=0}^\infty\phi_n(x_i)\phi_m(x_i)\rho(x_i)=\delta_{nm}d^2_n,
\label{ortog}
\end{equation}
where the weight function  $\rho(x)=1/[x C^2(x)]$ and
\begin{eqnarray}
 d_n^2=\frac{\Gamma(n)\Gamma(n+1)}{\beta^{2n}}.\nonumber
\end{eqnarray}
It follows from Eq.~(\ref{ortog}) that the wave functions $\phi_n(x)$ at
$n\neq m$ are orthogonal, provided that $0<\beta<1$ \cite{NikSysUvar}.
These wave functions are easy to modify in order that they also be
orthonormal.

The discrete mass spectrum $E_n$ corresponding to the wave functions
$\phi_n(x)$ in the form of Meixner polynomials \cite{NikSysUvar}
satisfies the simple equation
\begin{eqnarray}
 i(M^2-Q^2_{\rm out})=2n\sqrt{E_n^2-1}.
\end{eqnarray}
As a result, for the sought for mass spectrum we find
\begin{eqnarray}
 E_n^2=\left(\frac{m_{\rm out}}{M}\right)^2=1-\frac{(M^2-Q^2_{\rm out})^2}{4n^2}.
 \label{spektr}
\end{eqnarray}
The hydrogen-like mass spectrum found generalizes the result obtained
previously in \cite{BerKozKuzTkachPLB} and coincides with the result of
\cite{Berezin,BerezinPRD,BerBoyNerPRD}, where it was obtained by a different
method. A discrete mass spectrum takes place if $E_n<1$. If, alternatively,
$E_n>1$, then a continuous mass spectrum will take place \cite{BerezinPLB}.
Note that, in accordance with Eq.~(\ref{spektr}) for the mass spectrum, the
total gravitating mass of the metric $m_{\rm out}$ turns out to be lower
than the classical ``bare'' value of M due to the quantum corrections, as it
must be for a gravitationally bound system.

When $(M^2-Q^2_{\rm out})^2/4>1$, we can write a condition for the
applicability of the semiclassical approximation. Indeed, in this case,
there exists a minimum value of the quantum number, $n_{\min}= [(M^2-Q^2_{\rm out})/2]$, where ''$[\:\:]$`` denotes the integer part, for which the semi-classical approximation will definitely hold at $n\gg n_{\min}$. In the case, where the opposite inequality $(M^2-Q^2_{\rm out})^2/4<1$ holds, the semiclassical approximation is applicable at any n.

To write the solution of the original equation (\ref{BasicEquat}), the inverse change of variable $x\rightarrow ix$ should be made. As an example, let us write out the first several polynomials:
\begin{eqnarray}
 \phi(x)_{n=1}\!\! &=& \!\!ix\beta^{ix}\frac{\beta^2-1}{\beta^2}\sum_{k
 =-\infty}^{\infty}c_k\exp(-2\pi kx),\nonumber\\
 \phi(x)_{n=2}\!\! &=& \!\!ix\beta^{ix}\!\!\left[ix\!\!\left(\!1\!
 -\!\frac{1}{\beta^2}\!\right)^2\!\!
 +\!1\!-\!\frac{1}{\beta^4}\!\right]\!\sum_{k
 =-\infty}^{\infty}\!\!\!\!c_k\!\exp(\!-2\pi kx).\nonumber
 \label{solut1}
\end{eqnarray}
When the requirement that the wave functions be finite at infinity is fulfilled, the constants $c_k=0$ will be zero for all $k<0$. The remaining constants of integration can be easily found from the boundary conditions at the coordinate origin. This was done in \cite{Berezin,BerezinPRD} for $n=1$, and we will not give them here. Making the change $E=\cos\lambda$, the solution can be represented as $\phi(x)_n=P_n(x)\exp(-\lambda x)C(x)$, where $P_n(x)$ are some polynomials of degree $n$ \cite{Berezin}.

In the extreme case, where $m_{\rm out}=Q_{\rm out}$, the spectrum degenerates (does not depend on $n$) and $E = 1$. The absence of such an extreme state in the quantum spectrum is in complete agreement with the analogous result for the mass spectrum of a Reissner-Nordstr\"om black hole \cite{Barv02}. This result can be interpreted in terms of the postulated third law of thermodynamics for black holes, according to which the extreme state of a black hole is unattainable. At the quantum level, this means that the transitions (decays) of a black hole to the extreme state are impossible \cite{Das96,Das01,Medved}. Note that the formal solution of the wave equation (\ref{WaveRealEquat}) in the extreme case is
\begin{eqnarray}
 \phi(x)=C_1(x)+C_2(x)x,\nonumber
 \label{extremal}
\end{eqnarray}
where $C_1$ and $C_2$ are periodic functions with a period of 1. These periodic functions can be expanded into Fourier series to give
\begin{eqnarray}
 \phi(x)\!=\!\sum_{k=0}^{\infty}c_k\exp(-2\pi k x)
 +ix\sum_{k=0}^{\infty}d_k\exp(-2\pi kx),
\end{eqnarray}
where we made the inverse transformation $x\rightarrow ix$ and took into account the boundary condition at infinity. All of the unknown coefficients $c_k$ and $d_k$ can be found using the boundary conditions (\ref{boundcond}).

The case where $M^2\le Q^2_{\rm out}$ may be considered. In this case, $\sigma_{\rm in}$ can take on two values: $\sigma_{\rm in}=\pm 1$. For radii
$\rho<(Q^2_{\rm out}-M^2)/2m_{\rm out}<2m_{\rm out}$, the value of $\sigma_{\rm in}$ is negative and $\sigma_{\rm in}=-1$. It is easy to write out the wave equation and to solve it. The solution for the wave function will have the form $\hat{\phi}(x)=(-1)^x\phi(x)$, where $\phi(x)$ is solution (\ref{solution}). The corresponding mass spectrum will not change.

Let us now consider the more general case. where only $Q_{\rm in}=0$. To find the general solution of Eq.~(\ref{BasicEquat}), we will now make a different change of variable, namely, $x\rightarrow ix$. This transformation means the rotation through $\pi/2$ in the complex plane. Equation (\ref{BasicEquat}) will then be rewritten as
\begin{eqnarray}
 \phi(x-1)&+&\left(1+\frac{2m_{\rm in}Mi}{x}\right)\phi(x+1)
 \nonumber\\
 &+&i\frac{M^2-Q^2_{\rm out}}{x}\,\phi(x)=2E\phi(x).
\end{eqnarray}
The latter equation can also be expressed in terms of Meixner polynomials \cite{NikSysUvar} and its solution is
\begin{eqnarray}
 \phi_n(x)&=&C(x)
 \frac{\tilde{\beta}^x\Gamma(x+1)}{\tilde{\beta}^{2x+2n}\Gamma(\gamma+x)}\,
 \\ \nonumber
 &&\times\triangle^n\Bigg[\frac{\tilde{\beta}^{2x+2n}\Gamma(\gamma+x)}
 {\Gamma(x+1-n)}\Bigg],
\end{eqnarray}
where
\begin{equation}
 \tilde{\beta} = E-\sqrt{E^2-1},\; \gamma = i2m_{\rm in}M, \; C(x)=C(x+1),
 \label{tildebeta}
\end{equation}
and $C(x)$ is also a periodic function with a period of $1$. This function can be similarly expanded into a Fourier series. Let us write out the first two polynomials:
\begin{eqnarray}
 \phi(x)_{n=1} &=& [\tilde{\beta}^2(\gamma+x)-x]\tilde{\beta}^{x}C(x),
 \\
 \phi(x)_{n=2} &=& \tilde{\beta}^{x}[\tilde{\beta}^4(\gamma+x+1)(\gamma+x)
 \nonumber\\
 &&- 2\tilde{\beta}^2 x(\gamma+x)+x(x-1)]C(x).
\end{eqnarray}
If, as above, we set $x=x_i$ and $x_{i+1}=x_i+1$ and sum the wave functions with the weight
\begin{equation}
 \rho(x)=\frac{\Gamma(\gamma+x)}{\Gamma(1+x)\Gamma(\gamma)C^2(x)},
\end{equation}
then the above polynomials will be orthogonal \cite{NikSysUvar}:
\begin{eqnarray}
 \sum\limits_{x_i=0}^\infty\phi_n(x_i)\phi_m(x_i)\rho(x_i)=\delta_{nm}d^2_n,
\end{eqnarray}
where
\begin{equation}
d^2_n=\frac{n!\Gamma(n+\gamma)}{\tilde{\beta}^{2n}(1-
\tilde{\beta}^2)^\gamma\Gamma(\gamma)}.
\end{equation}
Making the inverse change of variable $x\rightarrow -ix$ we ultimately obtain the polynomials
\begin{eqnarray}
 \phi(x)_{n=1} &=& [\tilde{\beta}^2(\gamma-ix)+ix]\tilde{\beta}^{-ix}
 \sum_{k=-\infty}^{\infty}c_k\exp(-2\pi kx),
 \nonumber\\
 \phi(x)_{n=2} &=& \tilde{\beta}^{-ix}[\tilde{\beta}^4(\gamma\!
 -\!ix\!+\!1)(\gamma\!-\!ix)\!+2\tilde{\beta}^2 ix(\gamma\!-\!ix)
 \nonumber\\
 &+& ix(ix+1)]\sum_{k=-\infty}^{\infty}c_k\exp(-2\pi kx).
 \label{solut2}
\end{eqnarray}
It is easy to show that at $\gamma=0$, i.e., at $m_{\rm in}=0$, solution
(\ref{solut2}) of the wave equation will be transformed into solution (\ref{solut1}). The discrete mass spectrum is now specified by the equation
\begin{eqnarray}
 i(M^2-Q^2_{\rm out}+2m_{\rm in}M\tilde{\beta})=2n\sqrt{E_n^2-1}.
 \label{MnimSpektrMass}
\end{eqnarray}
The mass spectrum was found to be imaginary, because Hamiltonian (\ref{hamiltonian}) will no longer be Hermitian at $m_{\rm in}\neq0$. There exists one degenerate case where the spectrum is real: at $E_n^2=1$ and $M^2-Q^2_{out}+2m_{in}M\tilde{\beta}=0$. From these two equations we obtain the conditions for the parameters of the problem under which the mass spectrum is degenerate and real: $m_{out}=m_{in}+M$, $M= -m_{in}+ \sqrt{m_{in}^2+Q_{out}^2}$. This limiting case represents the transition from the discrete mass spectrum to the continuous one. To make the Hamiltonian Hermitian at $m_{\rm in}\neq0$, it is necessary to make the change of operators $A(x)B(p)\rightarrow\frac{1}{2}[A(x)B(p)+B^\ast(p^\ast)A^\ast(x)]$, which, in our case, corresponds to the change
\begin{eqnarray}
 \frac{1}{x}\exp\!\left(\!i\frac{\partial}{\partial x}\right)\!\to
 \!\frac{1}{2}\bigg[\frac{1}{x}\exp\!\left(\!i
 \frac{\partial}{\partial x}\right)\!
 +\exp\!\left(\!-i\frac{\partial}{\partial x}\right)\!\frac{1}{x}\bigg].
\end{eqnarray}
After this change of operators, the wave function for the Hermitian Hamiltonian is
\begin{eqnarray}
 \phi(x+i) &+& \phi(x-i)-m_{\rm in}M\bigg[\frac{\phi(x+i)}{x}
 +\frac{\phi(x-i)}{x-i}\bigg]
 \nonumber\\
 &-& \frac{M^2-Q^2_{\rm out}}{x}\,\phi(x)=2E\phi(x).
 \label{GenEquat1}
\end{eqnarray}
This equation is much more complex. Its solution and the energy spectrum must depend on two quantum numbers that define the mass spectrum of the inner, $m_{\rm in}$, and outer, $m_{\rm out}$, black holes. We managed to find only an approximate solution of this equation in the case where the shell mass $M$ was a small parameter of the problem, but the masses $m_{\rm out}$ and $m_{\rm in}$ were not small and of the same order of magnitude. For simplicity, we also assume that $Q_{\rm out}=0$. At large radii $\rho$, we see from Eq.~(\ref{energy}) that the difference of the black hole masses outside and inside the shell in the approximation under consideration is also a small quantity of the order of $M$. Consequently, the parameter of the problem $E$ is not small. Making, as above, the change of variable
$x\rightarrow ix$  in the wave equation (\ref{GenEquat1}), we will obtain an equation with a linear accuracy in small parameter $M$
\begin{eqnarray}
 &&\phi(x+1)+\phi(x-1) \\  \nonumber
 &&+im_{\rm in}M\bigg[\frac{\phi(x+1)}{x}+\frac{\phi(x-1)}{x-1}\bigg]
 =2E\phi(x).
\label{WavEquatMin}
\end{eqnarray}
We will seek a solution of Eq.~(\ref{WavEquatMin}) in the form $\phi(x)=\phi_0(x)+My(x)$. In the zeroth order in small parameter $M$, the wave equation will be reduced to a simple difference equation
\begin{eqnarray}
 \phi_0(x+1)+\phi_0(x-1)=2E\phi_0(x),
\end{eqnarray}
whose solution is
\begin{eqnarray}
 \phi_0(x)=C_1(x)\beta^{x}+C_2(x)\tilde{\beta}^x,
 \label{PeriodSol}
\end{eqnarray}
where the quantities $\beta$ and $\tilde\beta$ are defined by Eqs.~(\ref{beta}) and (\ref{tildebeta}), respectively, while the functions
$C_1(x)$ and $C_2(x)$ are periodic with a period of 1.

In the first order of smallness, the function y(x) satisfies the equation
\begin{eqnarray}
 &&y(x+1)-2Ey(x)+y(x-1)= \\
 &-&\!im_{\rm in}\!\left[C_1(x)\!\left(\frac{\beta^{x+1}\!}{x}\!
 +\!\frac{\beta^{x-1}\!}{x\!-\!1}\right)\!+
 \!C_2(x)\!\left(\!\frac{\tilde\beta^{x+1}\!}{x}\!
 +\!\frac{\tilde\beta^{x-1}\!}{x\!-\!1}\right)\!
 \right]\!,\nonumber
\end{eqnarray}
whose solution is
\begin{eqnarray}
 y(x)&=&\frac{im_{\rm in}}{2\sqrt{E^2\!-\!1}}\Bigg\{\!\!-\!C_1(x)
 \Big[\frac{\beta^{x+1}}{x} [2F(x,1,1\!+\!x,\beta^2)-\!1]\Bigg.
 \nonumber\\
 &&+\beta^{x-1}\left[\beta^2\Psi(x\!+\!1)+\Psi(x)\right]\Big]
 \nonumber\\
 &&+C_2(x)\Big[\frac{\tilde{\beta}^{x+1}}{x}
 [2F(x,1,1\!+\!x,\tilde{\beta}^2)-\!1]\Big.
 \nonumber\\
 &&+\Bigg.\tilde{\beta}^{x-1}
 [\tilde{\beta}^2\Psi(x\!+\!1)\!+\!\Psi(x)]\Big]\Bigg\}.
 \label{GenSol}
\end{eqnarray}
Here $F(x,1,1\!+\!x,\beta^2)$ and $F(x,1,1+x,\tilde{\beta}^2)$ are the
hypergeometric functions, while $\Psi(x)=\Gamma^{'}(x)/\Gamma(x)$ is the
logarithmic derivative of the gamma function.

Consider the physical meaning of this solution. First, solution (\ref{PeriodSol}) represents two waves traveling in opposite directions. This can be easily made sure if the solution is represented as
\begin{eqnarray}
 \Psi\sim\exp(-iEt\pm x\ln\beta),
\end{eqnarray}
which follows from the time-dependent Schrodinger equation
\begin{eqnarray}
 H\Psi=E\Psi=i\hbar\frac{\partial\Psi}{\partial t}.
\end{eqnarray}
The general solution (\ref{PeriodSol}) resembles the solution of the Schrodinger equation in a periodic potential (the Bloch function). However, this is only a qualitative similarity, because there are originally no periodic initial conditions in the problem under consideration. Since the solution obtained describes a continuous mass spectrum, the approximate solution of the original equation (\ref{WavEquatMin}) also has a continuous mass spectrum. We failed to find a solution with a discrete mass spectrum. As we see from the general solution (\ref{GenSol}), it is also periodic with a period of 1. It can also be seen that the wave structure of the equation remains unchanged. In contrast, the amplitude is now a function of the argument and can change according to the presented law due to the addition of a nonlinearity to the difference equation.

We considered here the simplest model of semiclassical quantization of a thin shell in the Reissner-Nordstr\"om metric. We used the hypothesis, based on the calculations of the mass spectrum for a simpler model, that the mass spectrum of the shell must depend on two quantum numbers in the case where the metric is the Schwarzschild one inside and outside the shell. This can be explained in a simpler language as follows. Consider the Schwarzschild black hole. There exist two space-time regions $R_{\pm}$ on the Carter-Penrose diagram for this metric. In accordance with the results of \cite{Much,BekMuch}, we will find for each of the regions $R_{\pm}$ that the black hole mass spectrum must depend on two quantum numbers.

We are grateful to V. A. Berezin for helpful discussions and valuable remarks. This work was supported by the Russian Foundation for Basic Research (project no. 100200635a), the Ministry of Education and Science of the Russian Federation (project NSh 3517.2010.2), and State contracts nos. 02.740.11.0244, 02.740.11.5092, and 02.740.11.0250. One of us (S.V. Chernov) also thanks the Dynasty Foundation of Noncommercial Programs for financial support.

\end{document}